\documentclass[10pt,letterpaper]{article}
\usepackage{amsmath}
\usepackage{amssymb}
\usepackage{cite}
\usepackage{color}
\usepackage{opex3}
\usepackage{txfonts}
\usepackage{epstopdf}
\usepackage{amssymb}
\usepackage{tipa}
\usepackage{graphicx}
\usepackage{txfonts}
\usepackage{amssymb}
\usepackage{extarrows}

\begin{document}
\title{Kerr-nonlinearity Enhanced Conventional Photon Blockade in Second-order-nonlinear System\\}
\author{Hongyu. Lin$^{1,2}$, Xiaoqian. Wang$^{1}$, Zhihai. Yao$^{1*}$ and Dandan Zou$^{3**}$\\}
\address{$^1$ Department of Physics, Changchun University of Science and Technology, Changchun - 130022, China \\
$^2$ College of Physics and electronic information, Baicheng Normal University, Baicheng - 137000, China\\
$^{3}$ School of Physics and Electronic Information, Shangrao Normal University, Shangrao - 334001, China\\}
\email{*yaozh@cust.edu.cn, **37949126@qq.com}

\begin{abstract}
In the recent publication [Phys. Rev. B 87, 235319 (2013)], the conventional photon blockade(CPB) was studied for the low frequency mode in a second-order nonlinear system. In this paper, we will study the CPB for the high frequency mode in a second-order nonlinear system with the Kerr nonlinearity filling in the low-frequency cavity.
By solving the master equation and calculating the zero-delay-time second order correlation function $g^{(2)}(0)$, strong photon antibunching can be obtained in the high frequency cavity. The optimal condition for strong antibunching is found by analyticcal culations and discussions of the optimal condition are presented. We find that the Kerr-nonlinearities can enhanced the CPB effect. In addition, this scheme is not sensitive to the reservoir temperature, which make the current system easier to implement experimentally.
\end{abstract}






\section{Introduction}
A single photon source is a device that can emit photons one by one, which plays a key role in the realization of quantum information processing and quantum communication~\cite{02,03}. To do this, the study of single photon source has become a hot research topic of modern quantum optics. In order to realize the single photon emission, photon blockade (PB)~\cite{04,05} is one of the main means, which is a phenomenon that a single photon in a nonlinear cavity will blocks the second photon entering~\cite{06,07}. In general, we think that if one system is driven by a classical light field, which can produce the sub-Poissonian light, if so, the system can realize PB. The phenomenon of PB can be judged by the zero-delay-time second-order correlation function $g^{(2)}(0)$. The $g^{(2)}(0)<1$ indicates the PB effect occurs. In theory, there are two physical mechanisms for implementing PB. One is found by Liew and Savona in a photon moleule system~\cite{08}, and the fundamental principle is destructive quantum interference between distinct driven-dissipative path ways~\cite{09,10,11}, which is called the unconventional photon blockade (UPB). It is show that strong antibunching can be realized in two coupled cavities with weak nonlinearities. Currently, many systems are predicted to exist the UPB effect, e.g., a optomechanical systems having the property of mutual coupling~\cite{12,13}, weak nonlinear photonic molecules modes~\cite{14}, the system with second-order nonlinearities~\cite{15,16,17,18}, optical cavities with a quantum dot~\cite{19,20}, bimodal coupled polaritonic cavities~\cite{21}, and the single-mode cavities coupled with three-order nonlinearities~\cite{aa,22,23,24}.

Besides the first physical mechanisms, the strong photon antibunching can also be obtained, which required the system have a large nonlinearities~\cite{j,n} with respect to the decay rate. The anticlustering effect is based on strong second-order nonlinearity is called conventional photon blockade (CPB). The CPB was first observed in an optical cavity coupled to a single trapped atom~\cite{26}, and it is realize the CPB mainly by filled the high value nonlinear medium in the cavity to form split energy level. Subsequently, CPB was found in many experiments, including quantum optomechanical systems~\cite{27,28,29,30,31}, semiconductor microcavity with second-order nonlinear properties~\cite{01,33}, and cavity quantum electro dynamics system~\cite{34,35,36}. In addition to its role in single photon source, PB has also been applied in single photon transistor~\cite{37}, interferometer~\cite{38}, quantum optical diode~\cite{33} and other devices. Recently, in Ref.~\cite{01}, the conventional photon blockade(CPB) was studied for the low frequency mode in a second-order nonlinear system.

In this work, we study the CPB based on a two-mode system, which consisting of two spatially overlapping single-mode semiconductor cavities, and we filled the weak Kerr-nonlinearities in the low-frequency cavity.
By analytic calculation, we obtain the optimal condition for strong antibunching, and the discussions of the optimal condition are presented. Numerical analysis and analytical analysis are obtained, respectively. Furthermore, a comprehensive comparison is made between numerical analysis and analytical analysis, we find the current system can realize CPB in the high frequency mode $b$, and the numerical solution with the analytical solution agree well each other. Meanwhile, through numerical analysis of the parameters and find that if you want to realize CPB, two conditions $g\gg\kappa$ and relatively weak drive strength must be satisfied at the same time. Comparing with the CPB in Ref.~[01], the contribution of the present scheme can be summarized as: (i) we study the CPB for the high frequency mode, while Ref.~[01] concerns the the low frequency mode. (ii) We add the Kerr nonlinearity to the system and find that the Kerr nonlinearity can enhance the photon antibunching. (iii) In current system, the CPB is immune to the the reservoir temperature, which makes the current system easier to implement experimentally.

The manuscript is organized as follows:
In Sec.~{\rm II}, we introduce the physical model.
In Sec.~{\rm III}, we illustrate the analytical conditions and physical mechanismfor the two-mode system .
In Sec.~{\rm IV}, we show the comparison of numerical and analytical solutions for the CPB.
Conclusions are given in Sec.~{\rm V}.

\section{Physical model}
\label{sec:2}
\begin{figure}[h]
\centering
\includegraphics[scale=0.50]{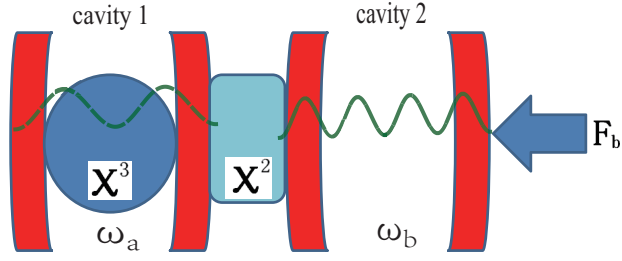}
\caption{Scheme of the system under investigation of
two coupled semiconductor microcavities, where the filled with Kerr-nonlinear materials cavity 1 is low-frequency cavitie, with frequency of $\omega_a$, cavity 2 is a high frequency cavity with frequency is $\omega_b$, and $\omega_b=2\omega_a$, and the two semiconductor microcavities are coupled by strong second-order nonlinearities, $F_b$ is the weak drives for mode $b$.} \label{fig1}
\end{figure}

The system consists of two single-mode cavities with frequencies $\omega_a$ and $\omega_b$, respectively. The cavitie 1 filled with three-order-nonlinear $\chi^{(3)}$ materials, the two cavities are coupled via strong two-order-nonlinear $\chi^{(2)}$ materials that mediates the conversion of the single-photon in cavity 2 into two-photon in cavity 1. We call the filled with Kerr-nonlinear materials cavity is a low-frequency cavity $a$, the another cavity is a high-frequency cavity $b$, the system model diagram is shown in Fig.~\ref{fig1}.

External weak drive is the key to realize PB. In this syetem we chose to drive on the high frequency cavity $b$. The driving frequency is $\omega_L$, and driving strength is $F_b$. So, this with Kerr-nonlinearity two modes system can be described by the following Hamiltonian~\cite{01,aa}
\begin{eqnarray}
\hat{H}&=&\omega_a\hat{a}^{\dag}\hat{a}\
+\omega_b\hat{b}^{\dag}\hat{b}
+g(\hat{b}^{\dag}\hat{a}^2+\hat{a}^{\dag2}\hat{b})+u\hat{a}^{\dag}\hat{a}^{\dag}\hat{a}\hat{a}
+F_b(\hat{b}^{\dag}e^{-i\omega_Lt}+\hat{b}e^{i\omega_Lt}),
\label{01}
\end{eqnarray}
The $\hat{a}(\hat{a}^{\dag})$ and $\hat{b}(\hat{b}^{\dag})$ denotes the annihilation (creation) operator of cavity $a$ and cavity $b$, respectively. the $g$ expresse the coefficient of second-order nonlinear interactions, which can be derived from the $\chi^{(2)}$ nonlinearity as~\cite{32}
\begin{eqnarray}
g=D\varepsilon_0(\frac{\omega_a}{2\varepsilon_0})\sqrt{\frac{\omega_b}{2\varepsilon_0}}\int d\textbf{r}\frac{\chi^{(2)}(\textbf{r})}{[\varepsilon(\textbf{r})]^{3/2}}\alpha_a^2(\textbf{r})\alpha_b(\textbf{r}).
\label{02}
\end{eqnarray}
The $u$ denote the strength of Kerr-nonlinear, which can be derived from the $\chi^{(3)}$ nonlinearity as~\cite{30}
\begin{eqnarray}
u=\frac{3\hbar^{2}\omega_a^{2}}{4\varepsilon_0}\int d\textbf{r}\frac{\chi^{(3)}(\textbf{r})}{[\varepsilon(\textbf{r})]^{2}}\alpha_a^4(\textbf{r}).
\label{03}
\end{eqnarray}
And the $\varepsilon_0$ is the vacuum permittivity, $\varepsilon_r$ is the relative permittivity, and $\alpha_a(\textbf{r})$ and $\alpha_b(\textbf{r})$ are the wave functions for mode $a$ and mode $b$, respectively.
In order to get an effective Hamiltonian, we use the rotating frame operator
$\hat{U}(t)=e^{i t(\omega_{l}\hat{a}^{\dag}\hat{a}+\omega_{l}\hat{b}^{\dag}\hat{b})}$ to operate on $\hat{H}$, which can lead to an effective Hamiltonian
$\hat{H}_{eff}=\hat{U}\hat{H}\hat{U}^{\dag}-i \hat{U}dU^{\dag}/dt$
as
\begin{eqnarray}
\hat{H}_{eff}&=&\Delta_a\hat{a}^{\dag}\hat{a}\
+\Delta_b\hat{b}^{\dag}\hat{b}+g(\hat{b}^{\dag}\hat{a}^2+\hat{a}^{\dag2}\hat{b})+u\hat{a}^{\dag}\hat{a}^{\dag}\hat{a}\hat{a}
+F_b(\hat{b}^{\dag}+\hat{b}),
\label{04}
\end{eqnarray}
The detuning of cavity $a$ and cavity $b$ can be expressed as $\Delta_a=\omega_a-2\omega_L$ and $\Delta_b=\omega_b-\omega_L$ .
The Fock-state basis of the system is denoted by $|m,n\rangle$ with the
number $m$ denoting the photon number in mode $a$, $n$ denoting the photon number in mode $b$. Under the weak driving limit, we restrict the system containing a single photon in the mode $b$. Hereafter, we select $|20\rangle$ and $|01\rangle$ to form a closed space, the Hamiltonian can be expand with $|2,0\rangle$ and $|0,1\rangle$, which can be described as a matrix form
\begin{eqnarray}
\tilde{H}=
\begin{bmatrix}
\Delta_a+2u & \sqrt{2}g\\
 \sqrt{2}g & \Delta_b
\end{bmatrix}, \label{05}
\end{eqnarray}
Because the system is operating under weak driving strength, so we have neglect the driving terms, and by analyzing the above matrix we can get two eigenfrequencies $\omega^{(1)}_{+}$ and $\omega^{(1)}_{-}$ in this hybrid system, which can be written as
\begin{eqnarray}
\omega^{(1)}_{\pm}&=&\frac{1}{2}[2\Delta_a + \Delta_b + 2u \pm \sqrt{
   8g^2 +4\Delta_a^2 +\Delta_b^2+4u^2-4\Delta_a\Delta_b-4\Delta_bu+8\Delta_au}].\nonumber\\
\label{06}
\end{eqnarray}
Here we set all of eigenvalues equal to zero, we can obtain the analytical conditions of CPB, which can be written as
\begin{eqnarray}
g=\pm\sqrt{\Delta_a\Delta_b+\Delta_bu}.
\label{07}
\end{eqnarray}
In order to realize CPB, the weak driving strength and the nonlinear coupling strength $g\gg\kappa$ must to be satisfied at the same time, the coupling strength $g\gg\kappa$ can ensure a large energy level splitting, and normally we choose to driving on the blocking pattern, which will facilitate the formation of anti-clustering of photons. Satisfy the above conditions, when other parameters are appropriate, the single photon blockade will occur in the cavity $b$.

\begin{figure}[h]
\centering
\includegraphics[scale=0.70]{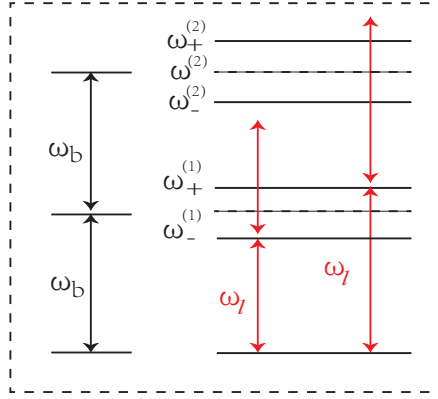}
\caption{(Color online) Schematic energy-level diagram explaining the occurrence of the single-photon blockade by the driving field satisfying the resonance condition when $\Delta_b=2\Delta_a$ is satisfied, where $\omega_{\pm}^{(1)}=\omega_b+u\pm \sqrt{2g^{2}+u^{2}}$ and $\omega_{\pm}^{(2)}=2\omega_b+2u\pm 2g$.} \label{fig2}
\end{figure}

The physical mechanism of CPB in the current system, which mixed with second-order nonlinear and Kerr-nonlinear system is analyzed next.
According to Eq.~(\ref{06}) we can get a energy level diagram, but it is complicated, so, for convenience, we can analyze the case that $\Delta_b=2\Delta_a$. Under this condition, the energy level diagram shown in Fig.~\ref{fig2}.
The physical mechanism of CPB is the anharmonic energy ladder. When Eq.~(\ref{06}) is met, the driving frequency $\omega_l$ equal to $\omega_{\pm}^{(1)}$, the single excitation resonance condition lead to the single-photon probability increasing dramatically. So the strong single-photon blockade can be triggered.

\section{The numerical results for the conventional photon blockade}
\label{sec:3}
In theory, to describe the statistical properties of photons we use the $g^{(2)}(0)$, the value of the $g^{(2)}(0)$ can be use to judged whether or not the single photon blockade happens, by solving the main equation we can get the expression for $g^{(2)}(0)$.
The dynamics of the density matrix $\hat{\rho}$ of the system is governed by
\begin{eqnarray}
\frac{\partial\hat{\rho}}{\partial t}&=&-i[\hat{H},\rho]+\frac{\kappa_a}{2}(\bar{n}_{th}+1)(2\hat{a}\hat{\rho}\hat{a}^\dag+\frac{1}{2}\hat{a}^\dag\hat{a}\hat{\rho}+\frac{1}{2}\hat{\rho}\hat{a}^\dag\hat{a})\nonumber\\
&&+\frac{\kappa_b}{2}(\bar{n}_{th}+1)(2\hat{b}\hat{\rho}\hat{b}^\dag+\frac{1}{2}\hat{b}^\dag\hat{b}\hat{\rho}+\frac{1}{2}\hat{\rho}\hat{b}^\dag\hat{b})\nonumber\\
&&+\frac{\kappa_a}{2}\bar{n}_{th}(2\hat{a}^\dag\hat{\rho}\hat{a}+\frac{1}{2}\hat{a}\hat{a}^\dag\hat{\rho}+\frac{1}{2}\hat{\rho}\hat{a}\hat{a}^\dag)\nonumber\\
&&+\frac{\kappa_b}{2}\bar{n}_{th}(2\hat{b}^\dag\hat{\rho}\hat{b}+\frac{1}{2}\hat{b}\hat{b}^\dag\hat{\rho}+\frac{1}{2}\hat{\rho}\hat{b}\hat{b}^\dag) ,\nonumber\\
\label{08}
\end{eqnarray}
The $\kappa_a$ and $\kappa_b$ denote the decay rates of cavities $a$ and $b$, respectively. Generally for the convenience of calculation and without loss of generality in the following discussion we set the detuning of these three cavities to satisfy the relationship of $\kappa_a=\kappa_b=\kappa$. $\bar{n}_{th}=\{\exp{[\hbar\omega/(\kappa_BT)-1]}\}^{-1}$ is the mean number of thermal
photons, $\kappa_B$ is the Boltzmann constant, and $T$ is the reservoir
temperature at thermal equilibrium.
In the current system, we only concern the  $g^{(2)}(0)$ in the steady state,
so, we just have to through setting $\partial\hat{\rho}/\partial t=0$ to solve the main equation for the steady-state density operator $\hat{\rho}_s$.
In this system, we will analyze the CPB in the high-frequency mode $b$, the statistic properties of photons will be described by the zero-delay-time correlation function, which is defined by
\begin{eqnarray}
g^{(2)}(0)=\frac{\langle
\hat{b}^{\dag}\hat{b}^{\dag}\hat{b}~\hat{b}\rangle}
{\langle \hat{b}^{\dag}\hat{b}\rangle^2}, \label{09}
\end{eqnarray}
The $g^{(2)}(0)$ can be calculated by solving the master
equations numerically, if the $g^{(2)}(0)<1$ indicates that the CPB occurs.

\section{Comparison of numerical and analytical solutions}
\label{sec:4}
\begin{figure}[h]
\centering
\includegraphics[scale=0.80]{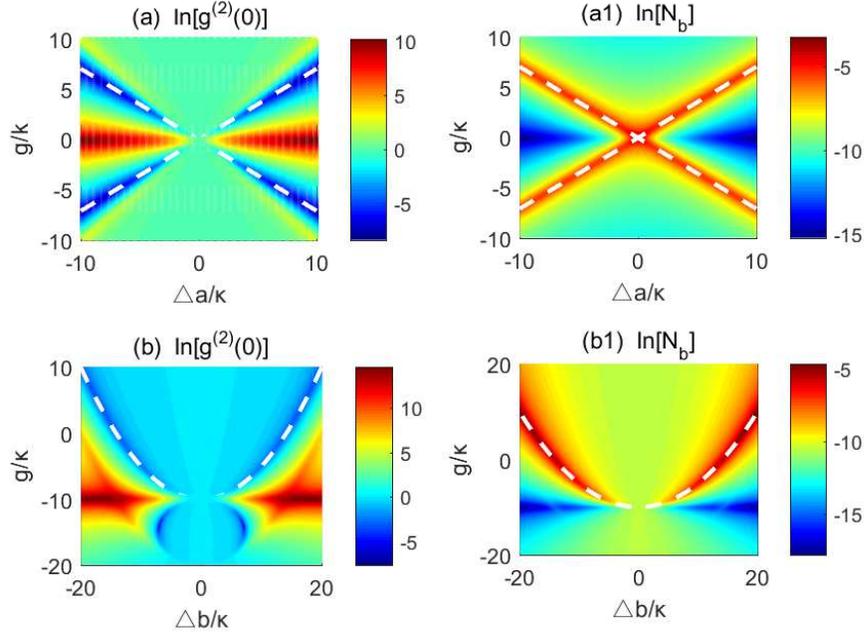}
\caption{(Color online) (a) and (a1)The Logarithmic plot of the $g^{(2)}(0)$ and average photon number $N_b$ as a function of $g/\kappa$ and $\Delta_a/\kappa$ for $F_a/\kappa=0.1$, $u/\kappa=0.5$, $\bar{n}_{th}=0$ and $\Delta_b/\kappa=2\Delta_a/\kappa$. (b) and (b1) The Logarithmic plot of the $g^{(2)}(0)$ and average photon number $N_b$ as a function of $g/\kappa$ and $\Delta_b/\kappa$ for $F_a/\kappa=0.1$, $u/\kappa=0.5$, $\bar{n}_{th}=0$ and $\Delta_a/\kappa=2$. In all the figures (a), (b), (a1) and (b1) the white dotted line denotes the optimal conditions of CPB shown in Eq.~(\ref{07}). All parameters are in units of $\kappa$ in this paper}
\label{fig3}
\end{figure}

Now, let us to study the CPB by the numerical simulation, and which compared it with optimal analytic condition show in Eqs.~(\ref{07}).
In order to represent the CPB effect, we plot of the $g^{(2)}(0)$ versus the system parameters to show the results, in a truncated Fock space. In the current system, the Hilbert spaces are truncated to six dimensions for the modes $a$ and $b$ respectively. And for convenience, we reset all parameters to units of dissipation rate $\kappa$, and the normalized analytic condition can be reformulated as
\begin{eqnarray}
g/\kappa=\pm\sqrt{(\Delta_a\Delta_b)/\kappa^2+(\Delta_bu)/\kappa^2}.
\label{10}
\end{eqnarray}
In Fig.~\ref{fig3} (a), (a1), (b) and (b1), we numerically study the CPB effect under the zero temperature ($\bar{n}_{th}=0$).
First of all, we logarithmic plot of the $g^{(2)}(0)$ as a function of $g/\kappa$ and $\Delta_a/\kappa$ for mode $b$, in Fig.~\ref{fig3}(a), the other parameters are $F_a/\kappa=0.1$, $u/\kappa=0.5$, $\bar{n}_{th}=0$ and $\Delta_b/\kappa=2\Delta_a/\kappa$. The numerical results show that, the CPB can occur in this system, where the valleys in $g^{(2)}(0)< 1$ corresponds to the strong photon antibunching, the shape of which similar to two diagonals in the region $g\gg \kappa$. The analytic solution is denoted by the white dotted line, which is agree well with numerical simulations. In Fig.~\ref{fig3}(b), we plot the $g^{(2)}(0)$ as a function of $g/\kappa$ and $\Delta_b/\kappa$ for mode $b$, and we set the $F_a/\kappa=0.1$, $u/\kappa=0.5$, $\bar{n}_{th}=0$ and $\Delta_a/\kappa=2$. The numerical results show that CPB effects can also occur, the shape of which looks like a quadratic function in the region $g\gg \kappa$. The analytic solution is represented by a white dotted line, and the same with Fig.~\ref{fig3}(a), the numerical solution is in good agreement with the analytical solution. In Fig.~\ref{fig3}(b), we fixed the value of $\Delta_a/\kappa=2$, the $\Delta_b/\kappa$ varies over a large area the CPB still occurs, this means that the CPB is insensitive to whether $\Delta_b/\kappa$ and $\Delta_a/\kappa$ are in a double relationship.
For further discussion the effect of CPB, in Fig.~\ref{fig3}(a1) we plot of the average photon number $N_b$ as a function of $g/\kappa$ and $\Delta_a/\kappa$, and in Fig.~\ref{fig3}(b1) we plot of the average photon number $N_b$ as a function of $g/\kappa$ and $\Delta_b/\kappa$, the coordinates and parameters are same with Fig.~\ref{fig3}(a) and (b), respectively. The brightness is defined as average photon number $N_b=\langle \hat{b}^{\dag}\hat{b}\rangle$, which can obtained be by numerically solving the master equation. Comparing with the Fig.~\ref{fig3}(a) and (b), the valleys with strong photon antibunching correspond to large average photon numbers, which illustrate that the current system has more mode $b$ photons produced in the CPB region, this further indicates that the system can realize single photon blocking. In general, under the weak driving condition, the single-photon probability is far larger than the two-photon probability, i.e., $P_{01}\gg P_{20}$, where $P_{01}$ denotes the probability that there are one photon in mode $b$, $P_{20}$ denotes the probability that there is two photons in mode $a$. However, the single excitation resonance with the Eq.~(\ref{10}) makes the $P_{01}\gg P_{20}$, the $P_{01}$ increasing dramatically due to the single excitation resonance, which makes the CPB occurs in mode $b$.

\begin{figure}[h]
\centering
\includegraphics[scale=0.60]{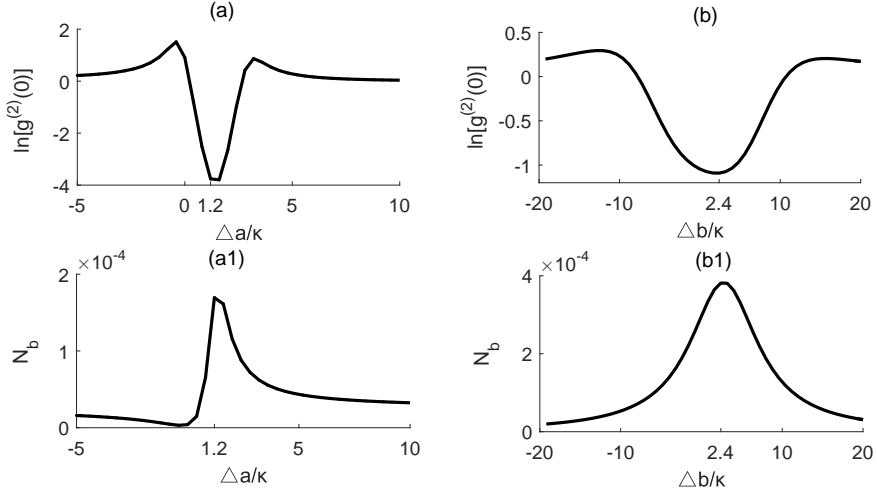}
\caption{(Color online) (a) and (a1)The Logarithmic plot of the $g^{(2)}(0)$ and average photon number $N_b$ as a function of $g^{(2)}(0)$ and $\Delta_a/\kappa$, with $u/\kappa=0.1$, $F_a/\kappa=0.1$, $g/\kappa=6$, $\bar{n}_{th}=0$ and $\Delta_b/\kappa=20$. (b) and (b1) The Logarithmic plot of the $g^{(2)}(0)$ and average photon number $N_b$ as a function of $g^{(2)}(0)$ and $\Delta_b/\kappa$ for $F_a/\kappa=0.1$, $u/\kappa=0.1$, $\bar{n}_{th}=0$ and $\Delta_a/\kappa=20$. Both of the four figures are numerical solution.}
\label{fig4}
\end{figure}

Next, in order to further compare the correlation between CPB and average photon number, in Fig.~\ref{fig4}(a) we plot the $g^{(2)}(0)$ as a function of $\Delta_a/\kappa$, with $u/\kappa=0.1$, $F_a/\kappa=0.1$, $g/\kappa=6$, $\bar{n}_{th}=0$ and $\Delta_b/\kappa=20$. Based on the existing parameters and the optimal condition shown in Eq.~(\ref{10}), the optimum blocking position should be in $\Delta_a/\kappa=1.7$, which is in good agreement with the numerical solution shown in the Fig.~\ref{fig4}(a).
In Fig.~\ref{fig4}(a1) we plot the average photon number $N_b$ as a function of $\Delta_a/\kappa$, the other parameters are same as Fig.~\ref{fig4}(a). Comparing the two diagrams, we find that the position of optimum blocking and the position of maximum average number of photons agree well with each other. In Fig.~\ref{fig4}(b), we plot the average photon number $N_b$ versus the nonlinear interaction strength $g/\kappa$, where $u/\kappa=0.1$, $F_a/\kappa=0.1$, $g/\kappa=5$, $\bar{n}_{th}=0$ and $\Delta_a/\kappa=20$, the optimum blocking position occurs at $\Delta_b/\kappa=2.4$, it is also in good agreement with the numerical results. in Fig.~\ref{fig4}(b1) we plot the average photon number $N_b$ as a function of $\Delta_a/\kappa$, the other parameters are the same as Fig.~\ref{fig4}(b), Same things are happening with Fig.~\ref{fig4}(a) and Fig.~\ref{fig4}(a1), the optimum CPB position and the point of maximum average number of photons also agree well with each other. Summarize the above results, the numerical solutions of the current system is completely corresponding to the analytical solutions, and the CPB occurrence regions is perfect consistent with the regions of average photon number increase. So, the current system can achieve single photon emission theoretically.

\begin{figure}[h]
\centering
\includegraphics[scale=0.60]{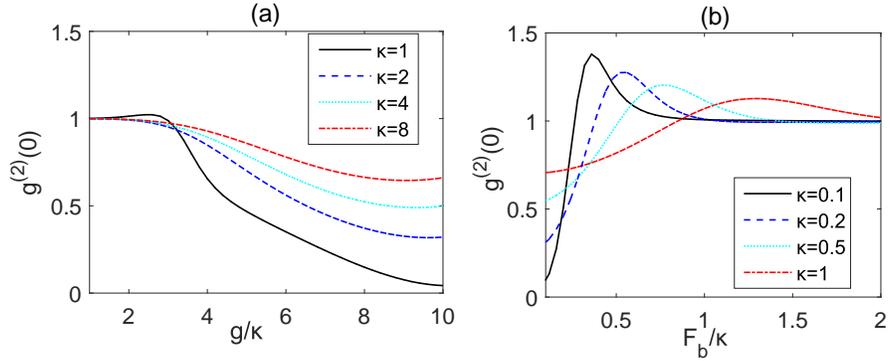}
\caption{(Color online) (a) Under the different dissipation rate $\kappa$, we plot the $g^{(2)}(0)$ vs the second-order nonlinear interaction strength $g/\kappa$, with $u/\kappa=2$, $F_a/\kappa=0.1$, $\bar{n}_{th}=0$, $\Delta_a/\kappa=2$ and $\Delta_b/\kappa=4$.
(b) Under the different dissipation rate $\kappa$, we plot the $g^{(2)}(0)$ vs the weak driving strength $F_b/\kappa$, with $u/\kappa=2$, $\bar{n}_{th}=0$, $\Delta_a/\kappa=2$, $\Delta_b/\kappa=4$ and $g/\kappa=4$.} \label{fig5}
\end{figure}

In the following, let us discussion the relationship between the dissipation rate $\kappa$ and the second-order nonlinear interaction strength $g/\kappa$, in Fig.~\ref{fig5}(a) we plot the $g^{(2)}(0)$ vs the $g/\kappa$, under the different dissipation rate $\kappa$. In the black solid line we set the dissipation rate $\kappa=1$, in the blue dotted line $\kappa=2$, in the green point line $\kappa=4$, and in the red dash dot line $\kappa=8$. CPB occurs in all four curves, and with the increase of the parameter $g/\kappa$ the intensity of CPB increases gradually. But in the red dash dot line the value of $\kappa=8$, it is close to the value of $g/\kappa$, and the CPB effect is already very weak. Therefore, conventional photon blockade must satisfy the condition of $g/\kappa\gg\kappa$. For further discussion the relationship between $\kappa$ and $g/\kappa$, at the same time focus on the effect of driving strength $F_b/\kappa$ in this system, in Fig.~\ref{fig5}(b) we plot the zero-delay-time second-order correlation functions $g^{(2)}(0)$ vs the weak driving strength $F_b/\kappa$, under the different dissipation rate $\kappa$. In the black solid line we set $\kappa=0.1$, in the blue dotted line $\kappa=0.2$, in the green point line $\kappa=0.5$, and in the red dash dot line $\kappa=1$, it is consistent with the results obtained in Fig.~\ref{fig5}(a), the increase of dissipation rate $\kappa$ will inhibit the intensity of CPB, and when the dissipation rate $\kappa$ is close to the value of $g/\kappa$, the intensity of CPB effect is obviously weakened. Not only that, but as we can see from the Fig.~\ref{fig5}(b), as the value of driving strength $F_b/\kappa$ increases, the value of $g^{(2)}(0)$ tends to be 1, the CPB phenomenons disappeared. Summarize the results of Fig.~\ref{fig5}(a) and Fig.~\ref{fig5}(b), we can find that to realize CPB, both conditions of $g\gg\kappa$ and weak drive strength $F_b/\kappa$ must be met at the same time, the calculated results are consistent with the theoretical prediction of the formation of CPB.

\begin{figure}[h]
\centering
\includegraphics[scale=0.50]{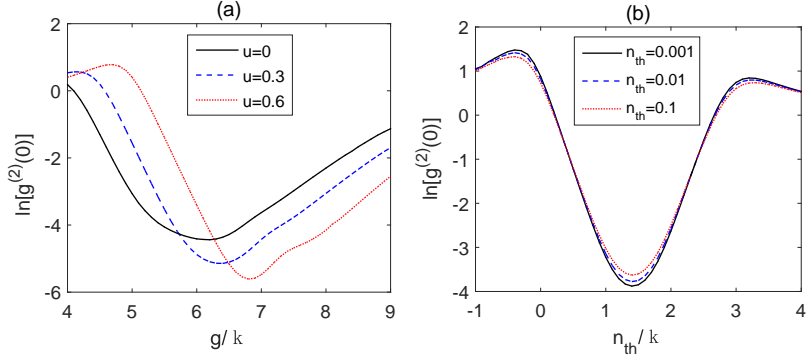}
\caption{(Color online) (a) Under the different third-order nonlinear coefficients $u/\kappa$, we plot the $g^{(2)}(0)$ vs the $g/\kappa$, with $F_b/\kappa=0.1$, $\bar{n}_{th}=0$, $\Delta_a/\kappa=3$ and $\Delta_b/\kappa=20$. In the black solid line we set $u/\kappa=0$, in the blue dotted line $u/\kappa=0.3$, in the red point line $u/\kappa=0.6$.
(b)Under the different number of thermal photons $\bar{n}_{th}$, we plot the $g^{(2)}(0)$ vs the detuning $\Delta_a/\kappa$, with $F_b/\kappa=0.1$, $u/\kappa=0.5$, $\Delta_b/\kappa=20$ and $g/\kappa=6$. In the black solid line we set $\bar{n}_{th}=0.001$, in the blue dotted line $\bar{n}_{th}=0.01$, in the red point line $\bar{n}_{th}=0.1$.} \label{fig6}
\end{figure}

Next, let us focus on the role of the third-order nonlinear coefficients $u/\kappa$ in this system, in Fig.~\ref{fig6}(a), we plot the $g^{(2)}(0)$ vs the $g/\kappa$, under the different values of $u/\kappa$, the other parameters in the Fig.~\ref{fig6}(a) are $F_b/\kappa=0.1$, $\bar{n}_{th}=0$, $\Delta_a/\kappa=3$ and $\Delta_b/\kappa=20$, in the black solid line we set $u/\kappa=0.1$, in the blue dotted line $u/\kappa=0.3$, and in the red point line $u/\kappa=0.6$. According to the given parameters, under the weak driving condition the position of the optimal blocking in good agreement with the analytical solution show in Eq.~(\ref{10}), the results are shown in Fig.~\ref{fig6}(a), as the value of $u/\kappa$ increases, the CPB strength obviously increases~\cite{23}.
In the previous research, we study the CPB effect with zero temperature. The last, we will investigate the CPB effect with nonzero temperature. in the Fig.~\ref{fig6}(b), we plot the $g^{(2)}(0)$ as a function of $\Delta_a/\kappa$, under the different number of thermal photons $\bar{n}_{th}$, where $F_b/\kappa=0.1$, $u/\kappa=0.1$, $\Delta_b/\kappa=20$ and $g/\kappa=5$, in the black solid line we set $\bar{n}_{th}=0.001$, in the blue dotted line $\bar{n}_{th}=0.01$, and in the red point line $\bar{n}_{th}=0.1$, the results show that the strongest CPB point appears on $\Delta_a/\kappa=1.3$, just as predicted. Moreover, when $\bar{n}_{th}$ changes in a large ranges, the PB does not change significantly, which indicate that this scheme is not sensitive to the change of the reservoir temperature, that make the system easier to implement experimentally.

\section{Conclusions}
\label{sec:5}
In summary, we have investigated the conventional photon blockade in a system consisting of two spatially overlapping single-mode semiconductor cavities, and which the low-frequency cavity filled with weak Kerr-nonlinearities, at the same time the low frequency cavity is coupled to the high frequency cavity by strong second-order nonlinearity. we consider the situation where the high frequency mode are driven simultaneously. We get strong photon antibunching conditions in the high frequency mode $b$. According to analytical simulation and numerical calculation, the two results are in good agreement. Under the different dissipation rate $\kappa$, we plot of the $g^{(2)}(0)$ vs the $g/\kappa$, and $g^{(2)}(0)$ vs the weak driving strength $F_b/\kappa$, the results show that to realize CPB, two conditions $g\gg\kappa$ and relatively weak drive strength must be satisfied at the same time, and the weak Kerr-nonlinearities can facilitate the implementation of CPB. Moreover, in this system the CPB is immune to the the reservoir temperature, which  makes the experimental realization of the present scheme easy.

\section*{Acknowledgment}
This work is supported by the National Natural Science Foundation of China with Grants No. 11647054, the Science and Technology Development Program of Jilin province, China with Grant No. 2018-0520165JH, the Jiangxi Education Department Fund under Grant No. GJJ180873.
\section*{Disclosures:} The authors declare no conflicts of interest.\\

\end{document}